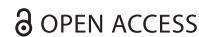
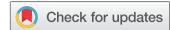

# What large-scale publication and citation data tell us about international research collaboration in Europe: changing national patterns in global contexts

Marek Kwiek

Center for Public Policy Studies, UNESCO Chair in Institutional Research and Higher Education Policy, University of Poznan, Poznan, Poland

**ABSTRACT**

This study analyzes the unprecedented growth of international research collaboration (IRC) in Europe during the period 2009–2018 in terms of co-authorship and citation distribution of globally indexed publications. The results reveal the dynamics of this change, as growing IRC moves European science systems away from institutional collaboration, with stable and strong national collaboration. Domestic output has remained flat. The growth in publications in major European systems is almost entirely attributable to internationally co-authored papers. A comparison of trends within the four complementary collaboration modes clearly reveals that the growth of European science is driven solely by internationally co-authored papers. With the emergence of global network science, which diminishes the role of national policies in IRC and foregrounds the role of scientists, the individual scientist's willingness to collaborate internationally is central to advancing IRC in Europe. Scientists collaborate internationally when it enhances their academic prestige, scientific recognition, and access to research funding, as indicated by the credibility cycle, prestige maximization, and global science models. The study encompassed 5.5 million Scopus-indexed articles, including 2.2 million involving international collaboration.

## 1. Introduction

International research collaboration (IRC) is central to contemporary higher education and science systems. Across Europe, the number and percentage of internationally co-authored publications continue to rise, as does the mean distance between collaborating scientists (Hoekman, Frenken, and Tijssen 2010). This internationalization is the defining feature of a new global geography of science (Olechnicka, Ploszaj, and Celinska-Janowicz 2019). However, following Caroline Wagner's theorizations on global and networked science, this paper contends that while the term *international science* connotes collaboration between nation-states, usually funded by governments, the emergent global science frees researchers to 'join forces to tackle common problems, regardless of where they are geographically based' (Wagner 2008, 31). It will be argued that the massive growth of collaborative science in Europe is not only a function of state and European Union promotion and funding but also, and perhaps more importantly, reflects individual scientists' pursuit of reputation and resources. In an era of increasing competition, the present analysis suggests that individual scientists' pursuit of collaboration with the best of their peers, regardless of location, is the primary driver of IRC growth in Europe (King 2011, 24).

Following earlier precedents in the literature, the concept of IRC is operationalized here as international co-authorship of scientific publications (Glänzel and Schubert 2001; Adams 2013) – that is, publications co-authored by scientists affiliated with institutions in at least two different countries as indicated in the article's byline. This aligns with the definition used in Scopus, the global dataset on which the study draws, exploring the internationalization of research as an outcome rather than a process (which is difficult to measure effectively) (Woldegiyorgis, Proctor, and de Wit 2018, 9). The study analyzes the unprecedented growth of IRC in Europe in terms of co-authorship and citation distribution among globally Scopus-indexed publications over the last decade (2009–2018). Particular attention is paid to the growing divide between the EU-15 and the EU-13 – that is, between the old and new European Union (EU) member states – in terms of IRC and its impacts.

Why has IRC increased more in Europe than elsewhere? First, Europe is a special case because policy has strongly promoted and funded IRC at both national and EU levels over the past two decades. Access to EU funding generally requires research partners from at least three countries, and both national and EU funding criteria have unambiguously favored internationalized principal investigators with large international collaboration networks and extended collaboration and mobility experience. For the period 2014–2021, the European Research Council (ERC) budget is 13.1 billion euro (König 2017, 42–59; see also Rodríguez-Navarro and Brito 2019 for an account of the ERC's limitations as the engine of European excellence). Under the 7th Framework Program for Research, 41.7 billion euro of the 50.5 billion euro budget for 2007–2013 was spent on about 26,000 projects, most involving international collaboration (Abbott et al. 2016, 309).

Secondly, IRC (both intra-European and beyond) has become a metric of excellence and quality within the European Research Area. In general, major European excellence initiatives of the last decade that offer additional and highly concentrated funding have also promoted IRC as their key goal. This accounts for IRC growth in Europe and its gradual emergence as one of the key criteria for academic promotion. In the globally unique European context (Fox, Realff, Rueda, and Morn 2017; König 2017; de Wit and Hunter 2017), IRC defines academic career prospects and determines individual and institutional access to national and European research funding. For that reason, the phenomenon of IRC in Europe merits special scholarly attention.

Third, in international collaborations at individual and institutional levels, 'excellence seeks excellence' (Adams 2013, 559); that is, scientists from top European universities predominantly seek to co-author with colleagues from top universities globally. High-performing institutions attract high-performing international collaborators, leading to highly cited joint papers. For instance, in 2009–2018, Oxford and Cambridge accounted for the largest number of international papers co-authored with the French CNRS, Harvard University, and Paris-Saclay University; ETH Zurich co-authored most international papers with CNRS, Paris-Saclay University and California Institute of Technology; and LMU Munich co-authored most international papers with CNRS, Harvard University and University College London. All of these are top performers in the global university rankings. The immense scale of IRC is revealed by the data; between 1996 and 2018, the percentage of Scopus-indexed publications (articles only) with authors from at least two European countries almost doubled (from 24.2% to 45.7%). The annual number of such articles grew almost four times – from 75,000 to 279,000 articles – to a total of 3.52 million articles published during that period. In 2018, almost half of the articles published in Europe and a third of those published in the OECD area (34.9%) involved international collaboration. In acknowledging these changing authorship practices and crediting those involved, it should be noted that these figures may reflect also an increasing number of authors per paper rather than merely a rising share of internationally co-authored papers. Finally, in accounting for this phenomenon, technological advances have had the same impact in Europe as elsewhere. Electronic communications make IRC faster and more efficient, and falling travel costs make the academic world smaller than ever before.

This study of changes within the European Union as a global leader in IRC addresses three research questions. (1) To what extent does IRC explain the massive growth in research output? (2) What are the major country-level collaboration networks as measured by publication quantity and (field-normalized) quality? (3) How does the citation premium for international collaboration differ by scientific field? Adopting a cross-national and cross-disciplinary perspective, key distinctions are drawn between (a) EU-15 and EU-13 and (b) the six major fields of research and development (FORD) used in OECD statistics. The literature review is followed by a brief description of data sources and methodology. Empirical results are then reported, followed by a discussion and conclusions.

## 2. Literature review

The topic of research internationalization has received much less scholarly attention than other aspects of internationalization such as teaching or cross-border mobility (Woldegiyorgis, Proctor, and de Wit 2018, 11). Perhaps the best answer to the more specific question of why IRC continues to grow is the simplest one: 'scientists collaborate because they benefit from doing so' (Olechnicka, Ploszaj, and Celinska-Janowicz 2019, 45). Scientists in Europe engage increasingly in international collaboration because they benefit more from this than from institutional or national collaboration. These patterns also reflect a drive by national governments and the European Commission (EC) to make IRC growth an explicit policy target (European Commission 2007, 2009; Lasthiotakis, Sigurdson, and Sá 2013).

### 2.1. IRC and the credibility cycle in academic careers

There is evidence that scientists increasingly seek IRC because it enhances academic recognition and provides better access to research funding (Jeong, Choi, and Kim 2014). The credibility cycle that enables European scientists to progress within their field (Latour and Woolgar 1986, 201–208) involves the conversion of prestigious articles into recognition, leading to grant-based funding, which is further converted into new data, arguments, and articles. IRC is a crucial component of this cycle. Internationally co-authored publications are the specifically European element of the publication-recognition link in this account of how academic careers develop. As a further European dimension, prestigious ERC grants and similar afford additional recognition (Van den Besselaar, Sandström, and Mom 2019). In competing for recognition, scientists vary in their individual predilection to collaborate and co-author internationally (Glänzel 2001, 69; Kwiek 2019b, 432–435): 'The more elite the scientist, the more likely it is that he or she will be an active member of the global invisible college' (Wagner 2008, 15) – that is, collaborating with colleagues in other countries. Scientists with an established reputation are more likely to collaborate internationally and so enter the global scientific elite. Highly visible and productive researchers work with those who are more likely to enhance their own productivity and credibility (Wagner, Park, and Leydesdorff 2015, 1616). At the same time, not surprisingly, members of these global elites 'might have performed better than others even without international collaboration' (Luukkonen, Persson, and Sivertsen 1992, 126).

In Europe, IRC is a prerequisite for establishing a successful individual career path. In European 'reputational work organizations' such as universities (Whitley 2000, 25), IRC is currently prioritized and funded as critically important in the struggle for resources and academic reputation. In Latour and Woolgar's (1986, 207) terms, IRC is widely reported to 'speed up the credibility cycle as a whole,' driving 'additional work and reputation in a virtuous circle' (Wagner and Leydesdorff 2005, 1616). In summary, IRC increases European scientists' chances of securing an academic position, moving faster up the career ladder, securing external funding for their research, and entering the global scientific elite.

As a consequence of how academic reward systems prioritize IRC and international mobility, hundreds of thousands of scientists travel by train and air across the relatively small, affluent, and scientifically advanced continent of Europe and co-publish at ever higher rates with European (and American) peers.

## 2.2. IRC and the prestige maximization model of universities

The growth of international collaboration in Europe can be also explained by the prestige maximization model of universities, which captures the changing dynamics of IRC and its financial and reputational implications. According to this model, which also captures the dynamics of global science, IRC is of increasing importance for individual and institutional success, and universities act principally as 'prestige maximizers' rather than 'profit maximizers' (Slaughter and Leslie 1997, 122–123; Melguizo and Strober 2007, 634). Focusing on individual prestige generation through publications, research grants, patents, and awards, the model posits a strong link between individual and institutional prestige: 'In maximizing their individual prestige, faculty members simultaneously maximize the prestige of their departments and institutions' (Melguizo and Strober 2007, 635). As prestige maximizers, universities and individual scientists must compete for critical resources and publication in high-impact journals – a key dimension of this competition (Slaughter and Leslie 1997, 114). In win-win cases, both the individual scientist and her institution maximize their prestige, which the global science community measures in terms of publications in elite journals, competitive research grants, and top academic awards (Kwiek 2018b, 2–3). In Europe over the last decade, prestige is increasingly maximized through internationally co-authored papers (although there are tensions related to the demise of traditional scholarly community norms still favor solo research in some fields) (Yemini 2019). The gradual transition from 'scientific nationalism' to the paradigm of 'global networked science' seems to parallel the increasing importance of individual ambition at the expense of broader national-level drivers of international collaboration.

## 2.3. IRC and the power of individual scientists

There is substantial support for the argument that the extent of IRC ultimately depends on the scientists themselves (Wagner and Leydesdorff 2005; King 2011; Royal Society 2011; Kato and Ando 2017; Wagner 2018), as faculty internationalization is seen to be shaped more by deeply ingrained individual values and predilections than by institutions and academic disciplines (Finkelstein, Walker, and Chen 2013) or governments and their agencies (Wagner 2018, x). In general, as research literature shows, IRC is influenced by academic discipline (with natural sciences being highly collaborative: Finkelstein and Sethi 2014, 235; Kyvik and Aksnes 2015, 1442), institutional type (with research universities being highly collaborative: Cummings and Finkelstein 2012, 86), and national reward structure (with internationalization traditionally being less important for promotion in central and eastern Europe: Dobbins and Kwiek 2017; Kwiek 2018c). IRC may also be related to gender, with female scientists possibly being more nationally and institutionally collaborative but less internationally collaborative than males (as in the Italian case in Abramo, D'Angelo, and Murgia 2013, 820; and as in the Polish case in Kwiek and Roszka 2020). The exceptions to this may be top research performers, who show no gender differences in collaboration patterns (Abramo, D'Angelo, and Di Costa 2019b, 416). However, as Aksnes, Piro, and Rørstad (2019, 770) found, the gender differences in the propensity to collaborate internationally in the case of Norwegian scientists are minor and not statistically significant. As a study of all German full professors in psychology found, male and female scientists may have different publication patterns: instead of submitting to competitive journals, female scientists may choose less competitive publishing venues (such as less prestigious book chapters: see Mayer and Rathmann 2018, 1675). In comparing productivity patterns in this specific sample, Mayer and Rathmann demonstrate that in the top 10% journals, there are considerably more men with a high publication output and considerably fewer men with a low publication output (Mayer and Rathmann 2018, 1676; see a study of the entire population of Quebec university professors in Larivière et al. 2011; and a study of 25,000 Polish university professors in Kwiek and Roszka 2020).

However, the decision to collaborate internationally in research is ultimately personal, and the concept of bottom-up 'self-organisation' (Wagner and Leydesdorff 2005, 1610; Wagner 2018, 84) is especially useful in understanding what drives collaborative global science. Increasingly, the motivation to internationalize comes from scientists themselves. European scientists tend to collaborate across national borders because they 'seek excellence' and want to work with the most outstanding scientists in their field (Royal Society 2011, 57); they seek 'resources and reputation' (Wagner and Leydesdorff 2005, 1616); and European academic reward structures incentivize them to exploit both collaboration and internationally co-authored publications to their own advantage (Glänzel 2001). To that extent, IRC is driven by an 'intrinsic motivation to succeed' and 'the motivation for better achievement' (Kato and Ando 2017, 2). As such, it is largely curiosity-driven and reflects 'the ambitions of individual scientists for reputation and recognition' (King 2011, 24). The traditional post-war 'governmental nationalism' in science co-exists with this global science, as scientists believe that their curiosity-driven (rather than state-driven) approach 'best serves their personal scientific ambitions' (King 2011, 361). While the role of national policy in directing scientific research diminishes, the influence of global networks seems to be growing (Wagner 2008, 24–25), extending and complementing the role of national systems (Wagner, Park, and Leydesdorff 2015, 11–12).

## 2.4. IRC and the global science model

Scientists – especially those in the elite layers of affluent systems – seem increasingly to act as free agents, carefully selecting research collaborators in what Wagner terms the general shift from 'national systems' to 'networked science' and moving freely within a global network (Wagner 2008, 25). According to Wagner, 'national prestige is not the factor that motivates scientists as they work in their laboratory benches and computers…. within social networks, scientists seek recognition for their work and their ideas' (Wagner 2008, 59). From this perspective, global science somehow goes on behind the backs of nation states; national systems fund institutions and scientists on the basis of merit but have little influence on collaboration patterns at global level (Wagner 2018, 177). The mechanisms of 'cumulative inequality' in global science mean that the rich (in reputation, citations, research funds, and personnel) get richer (King 2011, 368), and vertical stratification of the academic profession creates a divide between 'haves' and 'have-nots' (Wagner 2008, 1; see my monograph on the six major dimensions of social stratification in global science, Kwiek 2019a: academic salary stratification, academic power stratification, international research stratification, academic role stratification, and academic age stratification). Research is increasingly driven by collaboration between global elite groups (Adams 2013, 557); in Europe, Scopus collaboration data indicate that these are concentrated around London-Oxford-Cambridge, followed by Paris, Berlin-Munich, Stockholm-Uppsala and Lausanne-Zurich. These new inequalities are compounded by the value ascribed to knowledge produced in different countries and in different languages. As global science reproduces the global structure of center and periphery, core countries control knowledge flows and determine the rules of the academic game, imposing their research agendas and attracting talented scientists from the periphery (Olechnicka, Ploszaj, and Celinska-Janowicz 2019, 102–103).

Supported by new metrics for individual and institutional research evaluation and research assessment exercises across Europe, the global science model exerts a powerful 'pull' effect on scientists. As national ties weaken, the role of individual motivation seems to increase (Kato and Ando 2017), and individual scientists compete intensely within an 'economy of reputation,' involving 'battles over resources and priorities' (Whitley 2000, 26). In short, the growth of IRC in Europe is mainly an outcome of the rational choices of individual scientists seeking to maximize their own research output and impact (Hennemann and Liefner 2015, 345).

The dynamics of IRC in global science relate to the phenomenon of preferential attachment (Wagner 2008, 61–62; King 2011, 368) – that is, 'seeking to connect to someone already connected' (Wagner 2018, 76). A scientist's rising reputation and associated access to critical resources such as data, equipment, and funding means that 'other researchers are increasingly likely to want to form a link with her' (Wagner 2008, 61). Highly productive scientists attract similar individuals from elsewhere (King 2011, 368), and international networks form around these key people, who are highly attractive because they offer knowledge, resources, or both (Wagner 2018, 70). A large-scale data set of all Italian scientists indicates that productive scientists tend to collaborate more with international colleagues, and highly productive 'top performers' are much more internationalized than lower-performing colleagues (Abramo, D'Angelo, and Di Costa 2019a). Both large survey-based data (e.g. Kwiek 2016, 2018c) and smaller-scale, discipline-sensitive interview-based research (e.g. Yemini 2019) confirm that highly productive scientists are highly internationalized.

### 2.5. IRC: advantages and costs

The existing literature suggests that the advantages of IRC must be set against its costs, especially at national level (Wagner 2006). In particular, there is a risk that the academic peripheries may be unable, in the long run, to maintain their own research infrastructure, however critical for local purposes. At the individual level, a scientist's decision to engage in IRC must be viewed in the context of a trade-off between investment and expected outcomes. If it becomes overextended or too demanding, IRC can result in information overload, unclear responsibilities, and communication issues – collectively known as 'coordination costs' (Olechnicka, Ploszaj, and Celinska-Janowicz 2019, 111; Kwiek 2018a). Barriers to collaboration are compounded when the research involves international teams (e.g. Fox, Realff, Rueda, and Morn, 2017, 1294). Scientists make decisions about whether or not to collaborate internationally on the basis of available resources, the research environment, and trade-offs among alternative modes of collaboration (e.g. Jeong, Choi, and Kim 2014, 521).

## 3. Data sources and methodology

The data referred to here were retrieved between October 20–25, 2019 from Scopus, a database of abstracts and citations from the peer-reviewed literature, using its SciVal functionality. Scopus affords the best overview of the structure of world science, including most of the journals in the Thomson Reuters Web of Science (de Moya-Anegón et al. 2007; Lancho-Barrantes et al. 2012). Data for 24 EU member states from 2009 to 2018 were analyzed; the four remaining countries (Malta, Luxembourg, Cyprus, and Latvia) were removed from the analysis, as their total output was too small. All of the retrieved publication and citation data were aggregated to the six major fields of research and development used in OECD statistics: engineering and technologies, agricultural sciences, humanities, natural sciences, medical sciences and social sciences. The total number of included articles was 5.48 million, and the total number of citations was 87.48 million (2009–2018).

International collaboration was analyzed in the context of the three other collaboration types: institutional (all authors affiliated to the same institution); national (all authors affiliated to more than one institution within the same country); and single authorship (non-collaborative single-author outputs). This approach aligned with the structure of the Scopus and SciVal datasets; as the four collaboration types are complementary, publications can be divided into non-collaborative articles and those involving institutional, national, or international collaboration, and further aggregated into international collaborative articles and all others (referred to here as 'domestic articles').

## 4. Results

### 4.1. IRC, total national output, and system size in Europe

While standard input-output models of research evaluation were not employed here (see Godin 2007; Payumo et al. 2017), it is clear that lower levels of IRC – understood as the percentage of internationally co-authored publications – at the national level are positively correlated with lower levels of research expenditure in higher education systems in Europe. This correlation is confirmed in most EU-13 countries, where research underfunding is a dominant feature (leading some analysts to suggest a minimal investment threshold be imposed on all EU member states, as in Rodríguez-Navarro and Brito 2019, 14). However, the level of IRC in Europe is not generally correlated with national research output (defined as total number of articles 2009–2018) or number of research personnel (defined as researchers, full-time equivalent, higher education sector only, 2017). Plotting the percentage of internationally co-authored publications against system size in terms of publication numbers (Figure 1) and pool of academic researchers (Figure 2) reveals that correlations are negligible ($R^2 = 0.1$ and $R^2 = 0.06$, respectively). (In a regression model, R-squared values indicate the extent to which the variance of one variable explains the variance of a second; here, only 10% and 6% of the observed variation is explained by the model's inputs). In terms of publication output, the correlation is weak for top 100 nations ($R^2 = 0.21$), aligning with Lancho-Barrantes et al. (2012, 487). Bubble sizes in Figure 1 confirm that systems with low levels of IRC also have low field-weighted citation impact (FWCI) as defined by Scopus, as in the case of Croatia, Romania, and Poland (as well as EU-13 countries and China).

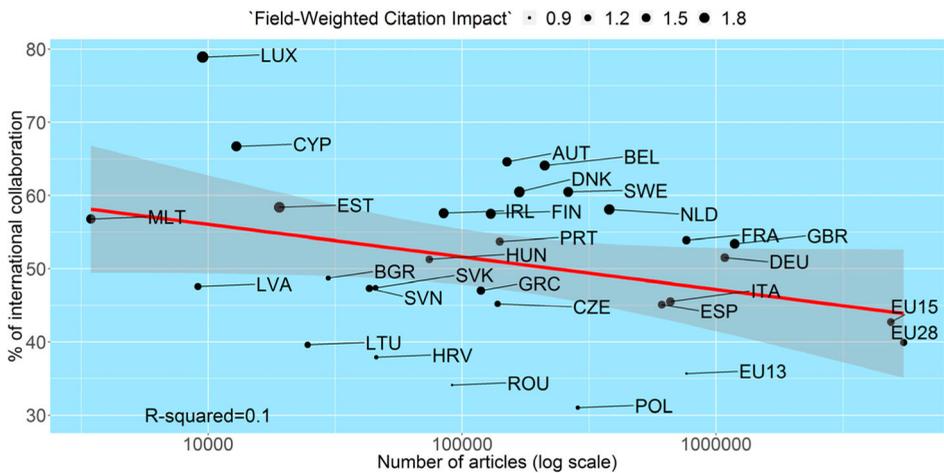

**Figure 1.** Correlation between total national output 2009–2018 (articles only; log number) and percentage share of publications in international collaboration, averaged for 2009–2018 (articles only); 95% confidence interval in gray; bubble size reflects average FWCI for the period.

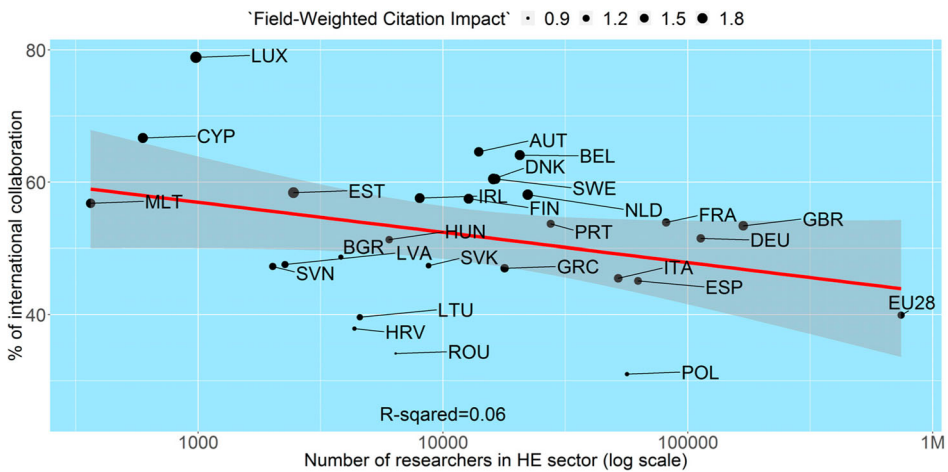

**Figure 2.** Correlation between national research personnel in the higher education sector at 2017 (FTE; category of researchers; log number) and average percentage share of publications in international collaboration 2009–2018 (articles only); 95% confidence interval in gray; bubble size reflects average FWCI for the period.

## 4.2. Changing collaboration patterns

Research collaboration trends can be analyzed in terms of changing percentage shares of the four major types of collaboration (international, national, institutional, none) and changing publication numbers over time. At the aggregated (all fields combined) level, European (as well as American and Chinese) data reveal clear growth in international collaboration, with stable national collaboration, and a substantial decline in the institutional category, supporting Adams' findings about the previous three decades (25 million Web of Science papers published between 1981 and 2012; Adams 2013). In all the European countries studied, IRC continues to grow, exceeding 50% in 2018 in all but three (Croatia, Poland, and Romania, all among the newest EU member states). Figure 3 (and Table 1) detail publication trends by collaboration type. In the natural sciences, traditionally characterized by high levels of IRC, there are even deeper changes, although the increase was much slower in EU-13 countries than in the EU-15. IRC was 60% or more in ten countries – that is, six out of ten articles originating from these countries had at least one international author. In terms of the share of total output, the leaders in research internationalization include eight small- and medium-sized systems (Austria, Belgium, Denmark, Sweden, Netherlands, Estonia, Finland, Ireland) and two large-sized systems (the United Kingdom and France). The group of internationalization leaders includes only one EU-13 country (Estonia).

National collaboration seems largely resistant to change; a decade of strong increase in IRC saw only a marginal decrease in national collaboration in most countries, with marginal increases in seven. National collaboration seems strongly embedded (possibly through state funding); based on strong intra-national scientific ties, it emerges as the most stable component of research collaboration over time. Across the EU-28, national collaboration decreased by only 0.5 percentage points during the study period, and in the USA, there was no change. However, institutional collaboration decreased in all of the countries studied, as did the share of single-authored papers (Table 1).

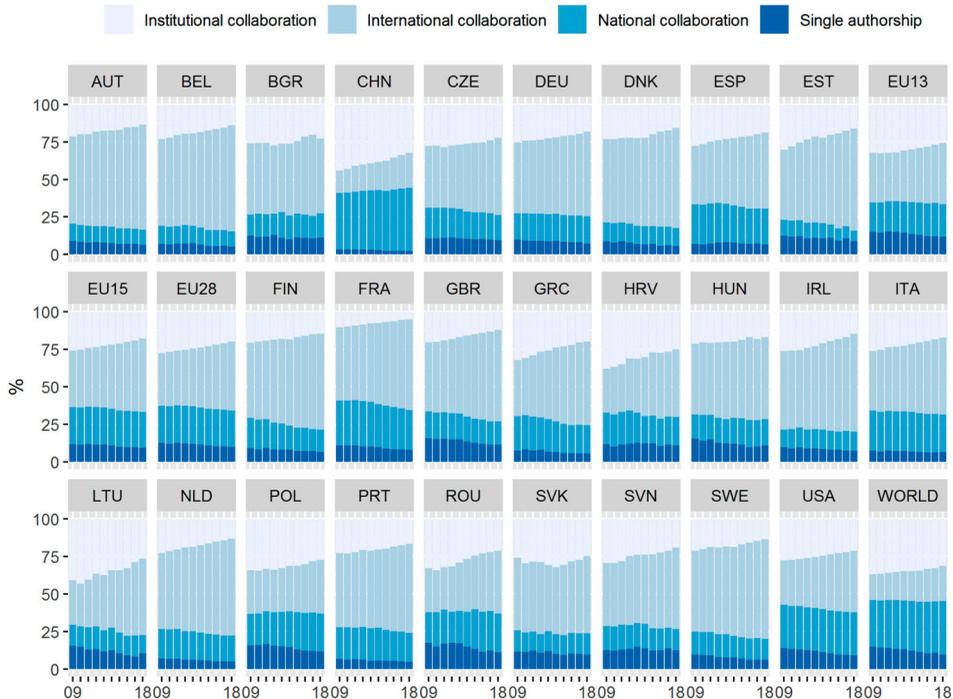

**Figure 3.** Increasing international collaboration at the expense of institutional collaboration, with stable national collaboration (for all fields of research and development combined): Europe as EU-28, EU-15, and EU-13 plus major EU-28 and comparator countries (articles only) 2009–2018 (%).

Table 1. Research collaboration trends over time: percentage of publications in EU-28 and comparator countries 2009 and 2018 (in descending order, by collaboration type, articles only, all fields of research and development combined) (%).

| International collaboration | | | | National collaboration | | | | Institutional collaboration | | | | Single authorship | | | |
| --- | --- | --- | --- | --- | --- | --- | --- | --- | --- | --- | --- | --- | --- | --- | --- |
| Country | 2009 | 2018 | Change in p.p. | Country | 2009 | 2018 | Change in p.p. | Country | 2009 | 2018 | Change in p.p. | Country | 2009 | 2018 | Change in p.p. |
| EST | 46.5 | 68.0 | 21.5 | ROU | 20.6 | 26.0 | 5.4 | SVK | 25.9 | 24.8 | −1.1 | SVN | 12.6 | 12.6 | 0.0 |
| LTU | 29.6 | 50.9 | 21.3 | World | 30.9 | 35.6 | 4.7 | BGR | 26.0 | 22.6 | −3.4 | ESP | 6.9 | 6.8 | −0.1 |
| GRC | 37.2 | 55.5 | 18.3 | CHN | 37.6 | 42.1 | 4.5 | HUN | 21.3 | 17.2 | −4.1 | HRV | 11.7 | 10.9 | −0.8 |
| HRV | 29.3 | 45.0 | 15.7 | POL | 21.1 | 25.4 | 4.3 | FRA | 10.5 | 5.3 | −5.2 | CZE | 10.6 | 9.7 | −0.9 |
| GBR | 45.7 | 60.9 | 15.2 | EU13 | 19.7 | 21.7 | 2.0 | World | 36.6 | 31.4 | −5.2 | CHN | 3.3 | 2.3 | −1.0 |
| FIN | 49.7 | 63.8 | 14.1 | BGR | 14.3 | 16.0 | 1.7 | CZE | 27.6 | 22.3 | −5.3 | ITA | 7.6 | 6.6 | −1.0 |
| NLD | 50.6 | 64.5 | 13.9 | HUN | 16.3 | 17.7 | 1.4 | PRT | 22.5 | 16.6 | −5.9 | BGR | 12.3 | 11.3 | −1.0 |
| IRL | 51.9 | 65.0 | 13.1 | IRL | 11.9 | 12.6 | 0.7 | FIN | 20.9 | 14.7 | −6.2 | BEL | 6.9 | 5.2 | −1.7 |
| BEL | 57.7 | 70.7 | 13.0 | DEU | 17.5 | 17.9 | 0.4 | USA | 27.6 | 21.2 | −6.4 | GRC | 7.5 | 5.8 | −1.7 |
| SWE | 53.6 | 66.6 | 13.0 | SVK | 14.1 | 14.1 | 0.0 | EU13 | 32.2 | 25.7 | −6.5 | NLD | 7.0 | 5.1 | −1.9 |
| SVN | 42.0 | 54.2 | 12.2 | USA | 28.8 | 28.8 | 0.0 | POL | 34.1 | 27.2 | −6.9 | PRT | 6.8 | 4.8 | −2.0 |
| ROU | 29.3 | 41.5 | 12.2 | EU28 | 24.7 | 24.2 | −0.5 | DEU | 25.3 | 18.2 | −7.1 | SVK | 11.8 | 9.8 | −2.0 |
| ESP | 38.9 | 50.7 | 11.8 | DNK | 12.7 | 12.0 | −0.7 | DNK | 23.1 | 15.6 | −7.5 | EU15 | 11.9 | 9.6 | −2.3 |
| ITA | 39.5 | 51.1 | 11.6 | EU15 | 24.7 | 23.7 | −1.0 | AUT | 21.4 | 13.6 | −7.8 | IRL | 9.8 | 7.5 | −2.3 |
| AUT | 58.2 | 69.8 | 11.6 | AUT | 11.4 | 10.1 | −1.3 | SWE | 21.2 | 13.4 | −7.8 | FIN | 9.1 | 6.7 | −2.4 |
| FRA | 48.7 | 60.2 | 11.5 | ITA | 26.6 | 25.0 | −1.6 | EU28 | 27.6 | 19.8 | −7.8 | DEU | 9.8 | 7.4 | −2.4 |
| EU15 | 37.4 | 48.7 | 11.3 | SWE | 15.5 | 13.7 | −1.8 | EU15 | 26.0 | 18.0 | −8.0 | EU28 | 12.6 | 10.2 | −2.4 |
| DNK | 55.5 | 66.8 | 11.3 | PRT | 21.1 | 19.2 | −1.9 | GBR | 20.6 | 12.2 | −8.4 | AUT | 9.1 | 6.5 | −2.6 |
| USA | 29.6 | 40.8 | 11.2 | SVN | 16.1 | 14.2 | −1.9 | ESP | 27.7 | 18.8 | −8.9 | FRA | 10.8 | 8.0 | −2.8 |
| EU28 | 35.1 | 45.7 | 10.6 | BEL | 12.3 | 10.2 | −2.1 | ITA | 26.2 | 17.3 | −8.9 | EU13 | 14.9 | 11.9 | −3.0 |
| CZE | 41.2 | 51.3 | 10.1 | HRV | 21.1 | 19.0 | −2.1 | BEL | 23.1 | 13.9 | −9.2 | DNK | 8.7 | 5.6 | −3.1 |
| PRT | 49.6 | 59.3 | 9.7 | LTU | 14.3 | 12.0 | −2.3 | NLD | 22.6 | 13.1 | −9.5 | SWE | 9.6 | 6.3 | −3.3 |
| DEU | 47.4 | 56.5 | 9.1 | GBR | 17.9 | 15.4 | −2.5 | SVN | 29.3 | 19.1 | −10.2 | EST | 12.5 | 8.8 | −3.7 |
| CHN | 14.9 | 23.4 | 8.5 | NLD | 19.8 | 17.2 | −2.6 | ROU | 32.7 | 21.3 | −11.4 | POL | 15.7 | 11.6 | −4.1 |
| HUN | 46.9 | 54.4 | 7.5 | ESP | 26.5 | 23.8 | −2.7 | IRL | 26.4 | 14.8 | −11.6 | GBR | 15.9 | 11.6 | −4.3 |
| EU13 | 33.2 | 40.7 | 7.5 | FRA | 30.0 | 26.5 | −3.5 | CHN | 44.1 | 32.3 | −11.8 | HUN | 15.5 | 10.8 | −4.7 |
| POL | 29.1 | 35.8 | 6.7 | EST | 10.7 | 7.1 | −3.6 | GRC | 32.2 | 19.8 | −12.4 | LTU | 15.4 | 10.7 | −4.7 |
| World | 17.5 | 23.4 | 5.9 | CZE | 20.5 | 16.7 | −3.8 | HRV | 38.0 | 25.2 | −12.8 | USA | 14.1 | 9.2 | −4.9 |
| SVK | 48.2 | 51.4 | 3.2 | GRC | 23.1 | 18.8 | −4.3 | EST | 30.3 | 16.1 | −14.2 | World | 15.0 | 9.7 | −5.3 |
| BGR | 47.4 | 50.0 | 2.6 | FIN | 20.3 | 14.8 | −5.5 | LTU | 40.8 | 26.4 | −14.4 | ROU | 17.4 | 11.2 | −6.2 |

The emergent dynamic of change is pervasive and clear; while national collaboration remains strong, dramatic growth in the internationalization of European research marks a shift away from institutional collaboration and single authorship. These processes are slower in the underperforming and resource-poor systems of Central and Eastern Europe (CEE), with powerful cross-disciplinary differences (see Kwiek 2020 on how "internationalists" in research differ from "locals" in research across academic fields, as well as across age, gender, academic seniority, working time distribution and academic role orientation).

### 4.3. IRC as the major driver of publication growth in Europe

The pervasive internationalization of European research is also reflected in the data on number of publications by collaboration type. National output can be divided into two categories: articles involving international collaboration and all others – that is, domestic articles, including both single-authored and national and institutional collaborations (see Adams 2013, 558). From this perspective, one dramatic finding is that the increase in annual output in 2009–2018 in such major European systems as the United Kingdom, France, the Netherlands, Finland, Belgium, Sweden, and Germany is entirely accounted for by international collaborations.

While domestic output in Europe remained almost flat during the study period, the number of internationally co-authored articles increased steadily (as was also the case in the USA). For instance, in a decade of rapidly expanding research output, the annual number of all domestic publications in the UK remained in the 54,000–59,000 bracket, with 54,104 publications in 2009 and almost exactly the same number in 2018 (54,121). In France, the equivalent range was 32,000–37,000 publications annually, with 34,432 in 2009 and 32,645 in 2018 (a 5.19% decrease). In Germany, there was a slight increase of 10.1% in the number of domestic publications. For EU-15 as a whole, the increase was 14.5%, and the US figure was similar (15.7%). However, the increases were much more substantial in EU-13 countries at 43.1%.

In the last decade, total annual research output has increased significantly (by 46.0% in EU-15 and by 30.9% in EU-13). However, the growth in publications in major European systems is almost entirely attributable to internationally co-authored papers. A comparison of trends within the four complementary collaboration modes clearly reveals that the growth of European science is driven solely by internationally co-authored papers. Figure 4 confirms this in the case of France, Germany, and the UK, the three largest European systems. The blue areas show the growth in numbers of international collaborative papers while the red line indicates the declining share of domestic publications. While the current power of research in western Europe resides in the growth of internationalization, the current weakness of research in CEE countries reflects their inability to keep pace with changes in the more affluent West, where the volume of internationally co-authored papers continues to increase.

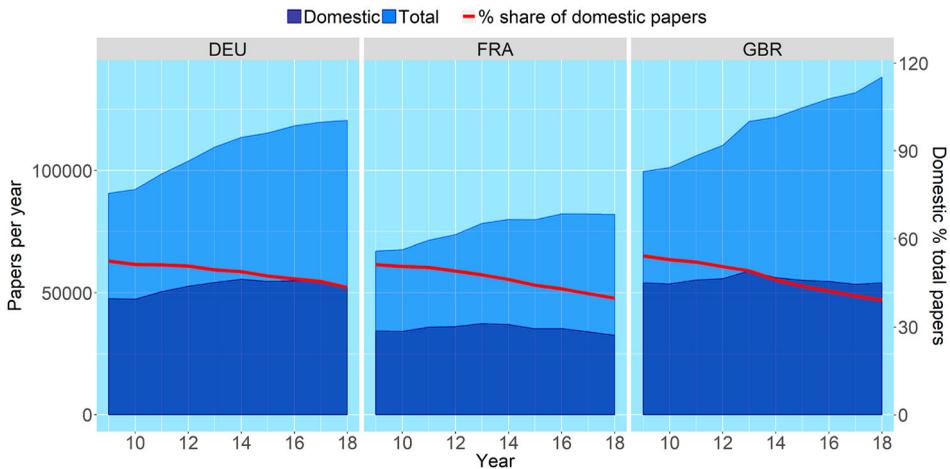

**Figure 4.** Total, domestic, and international collaborative publications for France, Germany, and the United Kingdom (2009–2018). All increase in total output is international collaboration; national collaboration remains flat in number, declining in percentage terms.

### 4.4. IRC and networks: major partnership countries

European countries' preferred research pairings differ significantly in terms of their global visibility (as operationalized by the Field-Weighted Citation Impact or FWCI of internationally co-authored publications). Field normalization of scientometric indicators avoids distortions caused by differing fields (Waltman and van Eck 2019, 282). As measured in Scopus, FWCI is the ratio of citations actually received to the expected world average for the subject field, publication type, and publication year.

For the majority of European countries, irrespective of the size of their science systems, the three most frequently collaborating partners are the USA, the UK, and Germany; for some others, preferred partners may also include France and Italy. Some collaboration patterns indicate that geographical, linguistic, and historical ties still matter; for example, Spain is the top collaboration partner for Portugal; Finland for Estonia; Germany for Austria and the Czech Republic; France for Romania; and the Czech Republic for Slovakia. The US remains the number one collaborating partner for most European countries, including the biggest knowledge producers (the UK, Germany, France, Italy, and Spain); these largest European knowledge producers are also the leaders in international collaboration (see Table 2, EU-28 countries only; and Table 3, EU-28 countries plus China and the USA; both tables in Data Appendices). In the top three ranks, however, FWCI is highest for the pairings of France and the Netherlands, Italy and the Netherlands, and Belgium and the United Kingdom. Within these top three pairs, internationally co-authored papers are cited 259–278% more than the world average for similar publications. If the US and China are included, the greatest number of internationally co-authored papers involves China and the United States, followed by the United Kingdom and the United States, Germany and the United States, and France and the United States. In short, the dominant feature of IRC in Western Europe is the predominance of collaboration with the US (country codes, see Table 4).

Table 2. Top 20 European collaboration partnerships (EU-28 countries only): most prolific pairs 2009–2018, sorted by number of co-authored publications (left) and field-weighted citation impact (FWCI) of co-authored publications (right).

| Rank | Partner Country 1 | Partner Country 2 | Publications 2009–2018 | FWCI | Rank | Partner Country 1 | Partner Country 2 | Publications 2009–2018 | FWCI |
|---|---|---|---|---|---|---|---|---|---|
| 1 | DEU | GBR | 134,073 | 2.91 | 1 | FRA | NLD | 40,961 | 3.78 |
| 2 | FRA | GBR | 95,833 | 3.12 | 2 | ITA | NLD | 39,187 | 3.71 |
| 3 | FRA | DEU | 95,447 | 2.96 | 3 | BEL | GBR | 38,121 | 3.59 |
| 4 | ITA | GBR | 90,551 | 3.00 | 4 | SWE | GBR | 44,967 | 3.46 |
| 5 | DEU | ITA | 80,744 | 3.10 | 5 | BEL | DEU | 35,663 | 3.46 |
| 6 | FRA | ITA | 76,693 | 2.94 | 6 | NLD | GBR | 75,417 | 3.33 |
| 7 | NLD | GBR | 75,417 | 3.33 | 7 | SWE | DEU | 41,046 | 3.27 |
| 8 | ESP | GBR | 72,460 | 2.99 | 8 | DEU | NLD | 72,336 | 3.17 |
| 9 | DEU | NLD | 72,336 | 3.17 | 9 | DEU | ESP | 62,027 | 3.15 |
| 10 | DEU | ESP | 62,027 | 3.15 | 10 | FRA | GBR | 95,833 | 3.12 |
| 11 | ITA | ESP | 60,153 | 3.01 | 11 | DEU | ITA | 80,744 | 3.10 |
| 12 | FRA | ESP | 58,851 | 3.09 | 12 | FRA | ESP | 58,851 | 3.09 |
| 13 | AUT | DEU | 52,290 | 2.49 | 13 | ITA | ESP | 60,153 | 3.01 |
| 14 | SWE | GBR | 44,967 | 3.46 | 14 | BEL | FRA | 40,976 | 3.01 |
| 15 | SWE | DEU | 41,046 | 3.27 | 15 | ITA | GBR | 90,551 | 3.00 |
| 16 | BEL | FRA | 40,976 | 3.01 | 16 | ESP | GBR | 72,460 | 2.99 |
| 17 | FRA | NLD | 40,961 | 3.78 | 17 | FRA | DEU | 95,447 | 2.96 |
| 18 | ITA | NLD | 39,187 | 3.71 | 18 | FRA | ITA | 76,693 | 2.94 |
| 19 | BEL | GBR | 38,121 | 3.59 | 19 | DEU | GBR | 134,073 | 2.91 |
| 20 | BEL | DEU | 35,663 | 3.46 | 20 | AUT | DEU | 52,290 | 2.49 |

Table 3. Top 20 collaboration partnerships, EU-28 countries plus China and USA: most prolific pairs 2009–2018, sorted by number of co-authored publications (left) and field-weighted citation impact (FWCI) (right).

| Rank | Partner Country 1 | Partner Country 2 | Publications 2009–2018 | FWCI | Rank | Partner Country 1 | Partner Country 2 | Publications 2009–2018 | FWCI |
|---|---|---|---|---|---|---|---|---|---|
| 1 | CHN | USA | 350,378 | 1.88 | 1 | NLD | USA | 89,626 | 3.33 |
| 2 | GBR | USA | 258,286 | 2.83 | 2 | NLD | GBR | 75,417 | 3.33 |
| 3 | DEU | USA | 216,945 | 2.69 | 3 | DEU | NLD | 72,336 | 3.17 |
| 4 | FRA | USA | 142,333 | 2.88 | 4 | DEU | ESP | 62,027 | 3.15 |
| 5 | DEU | GBR | 134,073 | 2.91 | 5 | FRA | GBR | 95,833 | 3.12 |
| 6 | ITA | USA | 127,454 | 2.80 | 6 | DEU | ITA | 80,744 | 3.10 |
| 7 | FRA | GBR | 95,833 | 3.12 | 7 | ITA | ESP | 60,153 | 3.01 |
| 8 | FRA | DEU | 95,447 | 2.96 | 8 | ESP | ITA | 60,153 | 3.01 |
| 9 | ESP | USA | 92,568 | 2.90 | 9 | GBR | ITA | 90,551 | 3.00 |
| 10 | ITA | GBR | 90,551 | 3.00 | 10 | ESP | GBR | 72,460 | 2.99 |
| 11 | NLD | USA | 89,626 | 3.33 | 11 | FRA | DEU | 95,447 | 2.96 |
| 12 | CHN | GBR | 82,782 | 2.27 | 12 | FRA | ITA | 76,693 | 2.94 |
| 13 | DEU | ITA | 80,744 | 3.10 | 13 | DEU | GBR | 134,073 | 2.91 |
| 14 | FRA | ITA | 76,693 | 2.94 | 14 | ESP | USA | 92,568 | 2.90 |
| 15 | NLD | GBR | 75,417 | 3.33 | 15 | FRA | USA | 142,333 | 2.88 |
| 16 | ESP | GBR | 72,460 | 2.99 | 16 | GBR | USA | 258,286 | 2.83 |
| 17 | DEU | NLD | 72,336 | 3.17 | 17 | ITA | USA | 127,454 | 2.80 |
| 18 | DEU | ESP | 62,027 | 3.15 | 18 | DEU | USA | 216,945 | 2.69 |
| 19 | ITA | ESP | 60,153 | 3.01 | 19 | CHN | GBR | 82,782 | 2.27 |
| 20 | ESP | ITA | 60,153 | 3.01 | 20 | CHN | USA | 350,378 | 1.88 |

Table 4. ISO 3-character codes by country.

| | | | |
|---|---|---|---|
| AUT | Austria | LVA | Latvia |
| BEL | Belgium | LTU | Lithuania |
| BGR | Bulgaria | LUX | Luxembourg |
| CHN | China | MLT | Malta |
| HRV | Croatia | NLD | Netherlands |
| CHE | Switzerland | POL | Poland |
| CYP | Cyprus | PRT | Portugal |
| CZE | Czech Republic | ROU | Romania |
| DNK | Denmark | SVK | Slovakia |
| EST | Estonia | SVN | Slovenia |
| FIN | Finland | ESP | Spain |
| FRA | France | SWE | Sweden |
| DEU | Germany | GBR | United Kingdom |
| GRC | Greece | USA | United States |
| HUN | Hungary | | |
| IRL | Ireland | | |
| ITA | Italy | | |

By way of example, Figure 5 examines Poland's and Germany's international collaboration partners more closely, plotting FWCI of publications involving each of the top 20 partners against the FWCI of all international publications involving that partner. Figure 5 shows how international collaboration increases FWCI of internationally co-authored papers for both Poland and Germany, as well as for their top 20 partners. There are clear mutual benefits in Poland's collaborations with Ukraine, as Poland's FWCI increases from 0.77 to 2.32 (horizontally) while Ukraine's FWCI increases two and a half times (vertically). Based on the citation premiums shown in Figure 5, all of these top 20 collaborations are win-win (quadrant 2).

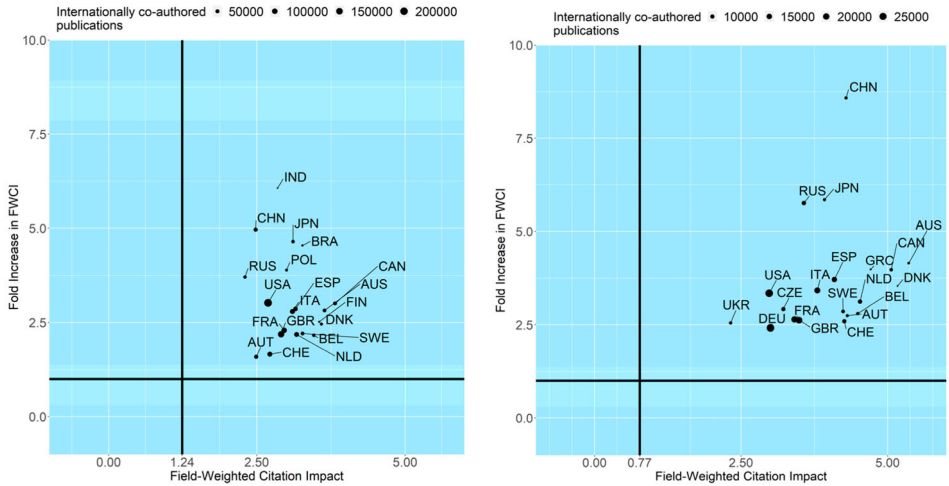

**Figure 5.** FWCI of publications involving international collaboration between Germany (left) and Poland (right) and their 20 largest partners. Horizontal lines indicate average FWCI (2009–2018) of all international collaborations among partner countries (=1); vertical lines indicate average FWCI (2009–2018, Poland and Germany) per international collaboration. Bubble size reflects number of joint internationally co-authored publications between 2009 and 2018 (all publication types, self-citations included).

### 4.5. Field differentiation of international collaboration premiums

As the extensive literature shows, internationally co-authored papers are cited more often for many reasons, not least because their authors are more likely to perform excellent research (Adams 2013, 559). This section examines the international collaboration premium (or superior citation returns) (Olechnicka, Ploszaj, and Celinska-Janowicz 2019, 100) in greater detail by field of research and development, relating the average number of citations of international or national co-authored publications to the benchmark of average institutional collaboration (100%) (see Kamalski and Plume 2013). Collaboration patterns by field are shown in Figure 6, revealing a clear distinction between old and new EU member states. Increases in citations of papers involving international collaboration are substantially higher for EU-13 than for EU-15 countries, especially in medical sciences (445% vs. 206% of baseline) and the humanities (348% vs. 148%), as well as for all fields combined (303% vs. 159%), reflecting global patterns (also shown). The smaller increases in the natural sciences may indicate that the citation premium for internationalization is lower in fields where collaboration has been the norm than in fields where it is expanding.

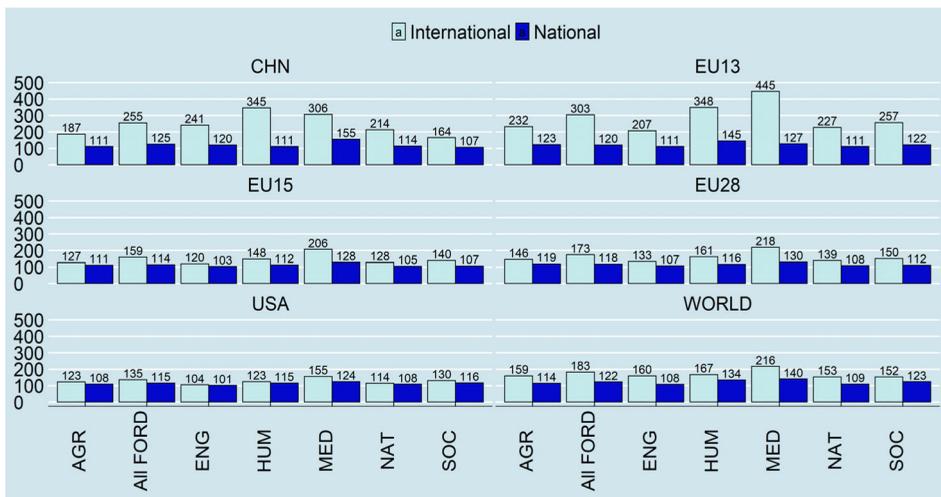

**Figure 6.** Citation premium for international and national collaboration, based on citation impact of institutional, national, and international collaboration, 2009–2018 (2009–2018 average, articles only, self-citations included) by field of research and development, country or aggregate country and increase over institutional collaboration (=100) (%).

At the same time, increases in citations of papers involving national collaboration are substantially lower in both EU-13 and EU-15 countries. For the USA, the increases are small (115% and 135%, respectively, for national and international collaboration for all fields combined). Increases are highest for medical sciences (155%) and lowest in engineering and technology (104%). In other words, international collaboration is most beneficial in EU-13 countries (and China) and least so in the USA, which aligns with previous studies (Wagner, Park, and Leydesdorff 2015, 15; Realff, Rueda, and Morn, 2017, 1303; Olechnicka, Ploszaj, and Celinska-Janowicz 2019, 92).

For all countries separately, however, the same analysis yields a much more nuanced picture of cross-national differences (Figure 7). The highest citation premium for international collaboration is found in EU-13 countries, with increases of up to 1,500% against the benchmark of 100% for institutional collaboration in the humanities in Bulgaria; in Romania, the increase is about 800%, and in Lithuania, about 700%. For social sciences, the increases exceed 500% in Bulgaria and 350% in Romania. In medical sciences, the increases are more than 700% in Bulgaria, 400–600% in the Czech Republic, Lithuania, Poland, Romania and Croatia, and 350% in Estonia and Hungary. In contrast, the average citation premiums for major EU-15 systems are much lower, with the exception of France and Spain (in humanities and medical sciences). While the fields of research with higher national relevance (either cultural, as in the humanities and social sciences, or practical, as in medical sciences) generally show lower levels of IRC in both EU-13 and EU-15 countries, the citation premium for international collaboration tends to be higher for these fields, especially in EU-13 countries. The fields with greater international validity, in which IRC tends to be more easily conducted and traditionally more prevalent, generally show a lower citation premium for international collaboration.

The striking EU-15/EU-13 divide is consistent with the idea that peripheries gain substantial international visibility through collaboration with centers (Glänzel and Schubert 2001; Wagner, Park, and

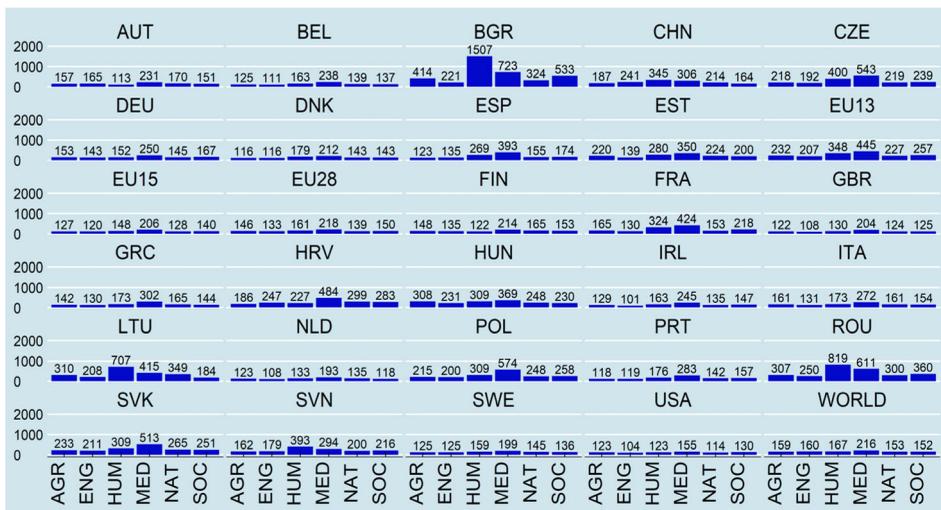

**Figure 7.** Citation premium for international collaboration, based on citation impact per institutional and international collaboration, 2009–2018 (2009–2018 average, articles only, self-citations included) by field of research and development, by country; increase over institutional collaboration (=100) (%).

Leydesdorff 2015). Interestingly, average citation premiums for national collaboration are not much different across European countries, with no observable EU-15/EU-13 divide.

Finally, international research collaboration can be analyzed in terms of the field-normalized impact of internationally co-authored papers on global science across countries. Using the standardized FWCI measure of publications by collaboration type, citations actually received are adjusted to the expected world average for the subject field, publication type, and publication year (through field normalization, Waltman and van Eck 2019, 281–300). SciVal provides the FWCI for national and international collaboration types, as well as for countries, institutions, disciplines, and individuals. An FWCI of 1.00 would indicate an exact match between a country's publications and the expected global average for similar publications (where FWCI for 'World' or the entire Scopus database is 1.00). An FWCI higher than 1.00 indicates that a country's publications are cited more (e.g. 2.11 means 111% more than the world average); conversely, an FWCI lower than 1.00 indicates that the country's publications have been cited less. For present purposes, this helps to explain the prestige of different European countries in terms of the extent to which their FWCI by collaboration type and field is above or below the world average over time.

As well as comparing citations intra-nationally (e.g. citations of all German papers written in international collaboration compared to the baseline of German papers written in institutional collaboration), citations actually received were compared cross-nationally in terms of FWCI – (for example, the actual global impact of German papers involving international collaboration was compared to the expected global impact of all such papers indexed in Scopus). In both cases, the analysis differentiated the six fields over time. This means that while the first approach compared national outputs intra-nationally, the second approach assessed prestige as the global impact of the various types of national output compared across countries and over time.

On comparing all collaboration types combined (international and national) for all six fields, the average FWCI for internationally co-authored papers for almost all EU-15 countries in all fields was (as expected) higher than the world average of 1.00 (i.e. those countries with horizontal lines above 1 in Figure 8). Publications involving international co-authors were cited more often than the global average, with the exception of Spain (medical sciences and social sciences) and Italy, France, and the United Kingdom (humanities). This finding confirms that domestic collaboration is more impactful in the humanities.

The impact of internationally co-authored papers from EU-13 countries is much lower and highly diversified by field. Poland and Romania are the only countries where impact is lower than the global average for all fields (for the whole decade in Poland and for almost the whole decade in Romania).

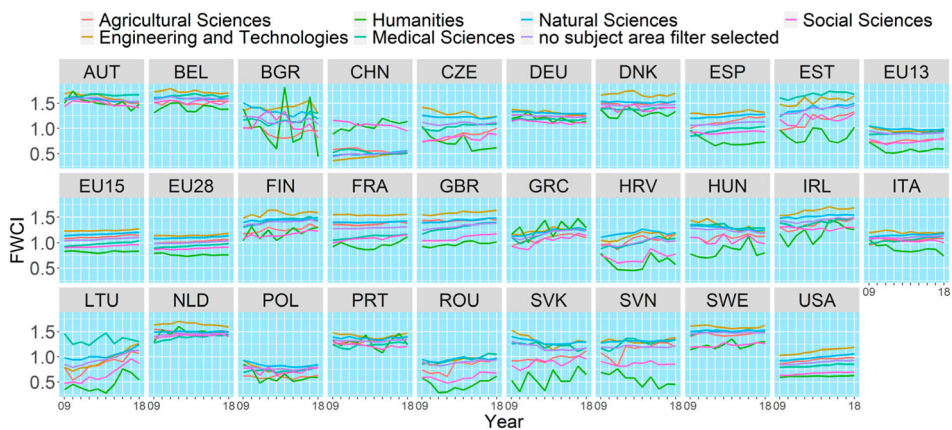

**Figure 8.** Field-weighted citation impact (FWCI) of internationally co-authored publications: articles only, self-citations included, by country and field of research and development, 2009–2018.

At this granular level, the most internationalized EU-13 country is Estonia, with only one field (humanities) below the global average. Consistent internationalization leaders include medical sciences in Lithuania and engineering and technology in the Czech Republic. Despite massive European funding and two waves of higher education reforms, Poland lags behind in all fields of research. Interestingly, the US and China fall into a group of countries where internationally co-authored papers in almost all fields (except for engineering and technologies in the US and except for humanities in China) are cited less often than the expected world average for this type of collaboration. In both cases, and especially in the case of the US, the central hub of the global collaboration network, this anomaly may be a function of size: the two global scientific powerhouses have large science sectors with plenty of internal possibilities for collaboration (Olechnicka, Ploszaj, and Celinska-Janowicz 2019, 79).

In contrast, nationally co-authored publications are cited less often than would be expected in almost all European countries (i.e. countries to the left of the vertical line in Figure 9), with EU-28, EU-15, China, and the US slightly above the global average. Papers involving national collaboration had a higher impact on global science than international collaborations in only five countries (those below the red dashed line), for different reasons: the global superpowers of China and the US, the European internationalization laggards of Poland and Romania, and France, where both nationally and internationally co-authored papers had a high impact. (Cross-disciplinary differences are not discussed here because of space constraints, see Kwiek 2015, 347–350.) At the aggregated level of all fields combined, the impact of internationally co-authored publications was above the expected field-weighted global average in the vast majority of European systems. The impact of papers involving national collaboration fell below this average (and are therefore located in quadrant 1). National collaboration produced globally impactful papers only in Portugal, Italy, Spain, and France (quadrant 2), as well as in the USA and China (quadrant 4).

## 5. Discussion and conclusions

In quantitative terms, Europe is clearly the global IRC leader. The total number of articles involving international collaboration during the period studied (2009–2018) was about 2.2 million in the EU-

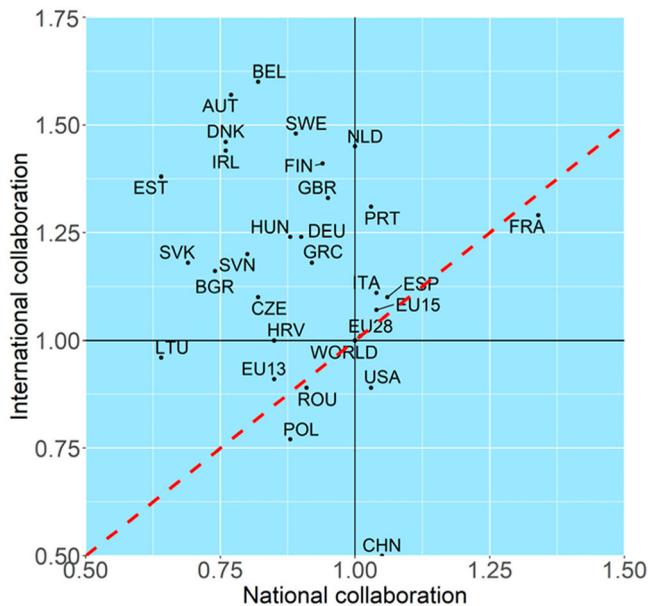

**Figure 9.** Field-weighted citation impact (FWCI) by publication type (internationally co-authored, nationally co-authored, articles only, self-citations included), average for 2009–2018, all fields of research and development combined.

28 as compared to about 1.4 million in the US and about 0.7 million in China. Globally, about 490,000 articles involving international collaboration were published in 2018, of which 57.4% involved co-authors from EU-28 countries. In the EU-28 45.7% of articles involved international collaboration; in the US, the rate was 40.8%. In ten countries, six out of ten articles had at least one international author. The research internationalization leaders include two large-sized systems (the United Kingdom and France) and eight small- and medium-sized systems. However, IRC in Europe is not dependent on total national research output or on the number of research personnel. (When IRC was plotted against publication numbers and researcher numbers, correlations proved negligible). At the same time, Europe's future as a global scientific powerhouse has been called into question on qualitative grounds. This is due to the lower than expected numbers of breakthrough papers (Rodríguez-Navarro and Brito 2019) or papers leading to breakthrough achievements among the most highly cited European publications. As the authors conclude, '[I]f the EU genuinely wants to recover its past status as a global scientific powerhouse … then strict measures should be taken by all EU countries regarding research' (Rodríguez-Navarro and Brito 2019, 15).

International collaboration in Europe is closely linked with European integration and EU research funding. The rationale behind supporting IRC at the EU level is reported to be twofold: '(1) to enhance Europe's scientific excellence, and (2) to spur European integration' (Olechnicka, Ploszaj, and Celinska-Janowicz 2019, 137). Collaboration has steadily increased in the last two decades as a result of the EU funding incentives. A report on the effectiveness of fostering Europe's international competitiveness in science shows that the share of internationally collaborative papers published by participants in the 7th Framework Program for Research and Technological Development (2007–2013), the predecessor of the current Horizon 2020 program, increased by 11.5–11.9 percentage points in major funding streams (European Commission 2015, 26). Collaborations begun with generous EU funding often continue after funding ends. Importantly, the Schengen Area comprises 26 European states that have officially abolished all passport control at their mutual borders; a single jurisdiction for international travel purposes, combined with massive EU funding, clearly supports international research collaboration (and research-linked international mobility, not analyzed here). However, EU funding is reported to only enable scientists to establish new international research collaborations rather than to create effective collaborations (in terms of productivity) (Defazio, Lockett, and Wright 2009, 304). A single market for research is developing within the EU: a study comparing intra-EU and extra-EU co-publication patterns shows a combination of Europeanization and internationalization trends (Mattsson et al. 2008, 573). International collaboration also includes bi-regional collaboration (such as EU-28 co-publishing with Latin America), with the US's powerful role as the main collaborating partner for almost all countries involved in such collaborations (Belli and Baltà 2019, 1465).

The present study shows that the dramatic growth of internationalization is moving European systems away from institutional collaboration and single authorship while national collaboration remains strong. With similar but slower processes in underperforming CEE countries, a decade of change in Europe shows domestic output remaining flat, with internationally co-authored articles increasing steadily. While total research output has increased dramatically (by 46.0% in EU-15 and by 30.9% in EU-13), this growth is attributable almost entirely to internationally co-authored papers. The dominant feature of IRC in Europe is the strength of collaboration with the US; the United Kingdom, Germany, and France collaborate more intensively with the US than any European country collaborates with any other European country (Table 3). Nevertheless, collaboration patterns indicate that geographic, linguistic, and historical ties remain strong. In general, IRC pays off in terms of citation premium in European systems; all collaborations with top 20 partners are win-win, increasing citation rates for both partners.

The present analysis applied two approaches. First, citations actually received by papers involving international collaboration were compared intra-nationally with the baseline of citations of papers involving institutional collaboration. Secondly, using the FWCI parameter, citations actually received were compared cross-nationally and to the global baseline value of 1.0. At the level of all fields combined, the field-normalized citation impact of internationally co-authored papers in almost all European systems was above the global average.

One major finding relates to the widening EU-15/EU-13 gap in research internationalization. This is a consequence of the long-term isolation of CEE countries from global science networks, along with severe underfunding of research systems. IRC is expensive and requires a basic threshold of public research funding, which has not been reached in CEE countries over the last three decades (see Dobbins and Kwiek 2017). The dominance of national publication patterns contributes further to this gap, with little institutional pressure on academics to publish internationally or in international collaboration for career advancement as compared to EU-15 countries.

With the emergence of the global network science, the role of national policy has diminished while individual scientists take center stage (Wagner, Park, and Leydesdorff 2015, 15). In Europe, and especially in CEE countries, the individual scientist's willingness to collaborate internationally is the key to advancing IRC. According to Eurostat, there were 743,364 FTE researchers in the higher education sector able to participate in IRC in 2017, often with generous EU funding. Ultimately, abstract statistical constructs relating research internationalization to 'EU-15,' 'countries,' and 'institutions' refer to aggregates of individual scientists who collaborate and publish internationally. To understand the future of the research internationalization agenda in Europe, it is essential to understand IRC success at this individual level, and how individual scientists make decisions about their involvement in international research (see substantial gender disparities in IRC in Kwiek and Roszka 2020). Although these decisions are strongly constrained and reflect 'the power of scientific networks and scientific standards to influence such choice making' (King 2011, 366), the choices that scientists make are also individual, autonomous, and decentralized. To that extent, IRC is 'essentially a bottom-up activity,' regardless of national or institutional strategies (Woldegiyorgis, Proctor, and de Wit 2018, 12), multinational programs, or memoranda of understanding (Adams 2013, 560). The individual scientist holds the key to IRC because she decides whether and with whom to collaborate and co-author, based on the reputation, resources, research interests, and general attractiveness of the potential research partner (Wagner 2018).

From a policy perspective, a fine-grained, cross-disciplinary analysis of science publishing trends across Europe can identify fields that are more or less positively affected by international collaboration. Detailed field-level and institution-level studies are especially relevant for EU-13 countries, which stand to benefit most from international collaboration and enhanced visibility. At a higher level of granularity, the Scopus data on All Science Journal Classification (ASJC) disciplines can be combined with data for individual universities and departments to identify fields of research and ASJC disciplines with very high or very low citation premiums as a basis for internationalization planning.

Finally, as European scientists seem to collaborate and co-author internationally in pursuit of academic prestige, scientific recognition, and access to research funding, it seems clear that individual choices are motivated by existing reward structures, including funding regimes and research policies, that prioritize research internationalization. The success of that internationalization owes to the vast network of collaborating scientists, funded by national governments and the European Union. As scientists leave behind the age of 'scientific nationalism' and enter the era of global science, their decisions to internationalize are more autonomous than ever before.

## Acknowledgements

The author gratefully acknowledges the support of the Ministry of Science and Higher Education through its Dialogue grant 0022/DLG/2019/10 (RESEARCH UNIVERSITIES). The support of Dr. Wojciech Roszka is also gratefully acknowledged.

## Disclosure statement

No potential conflict of interest was reported by the author(s).

## ORCID

*Marek Kwiek*  http://orcid.org/0000-0001-7953-1063

## References

Abbott, A., D. Butler, E. Gibney, Q. Schiermeier, and R. Van Noorden. 2016. "Boon or Burden: What Has the EU Ever Done for Science?" *Nature* 534: 307–9.

Abramo, G., C. A. D'Angelo, and F. Di Costa. 2019a. "The Collaboration Behavior of Top Scientists." *Scientometrics* 118 (1): 215–32.

Abramo, G., C. A. D'Angelo, and F. Di Costa. 2019b. "A Gender Analysis of Top Scientists' Collaboration Behavior: Evidence from Italy." *Scientometrics* 120: 405–18.

Abramo, G., C. A. D'Angelo, and G. Murgia. 2013. "Gender Differences in Research Collaboration." *Journal of Informetrics* 7: 811–22.

Adams, J. 2013. "The Fourth Age of Research." *Nature* 497: 557–60.

Aksnes, D. W., F. N. Piro, and K. Rørstad. 2019. "Gender Gaps in International Research Collaboration: A Bibliometric Approach." *Scientometrics* 120: 747–74.

Belli, S., and J. Baltà. 2019. "Stocktaking Scientific Publication on Bi-regional Collaboration Between Europe 28 and Latin America and the Caribbean." *Scientometrics* 121 (3): 1447–80.

Cummings, W. K., and M. J. Finkelstein. 2012. *Scholars in the Changing American Academy. New Contexts, New Rules and New Roles*. Dordrecht: Springer.

Defazio, D., A. Lockett, and M. Wright. 2009. "Funding Incentives, Collaborative Dynamics and Scientific Productivity: Evidence from the EU Framework Program." *Research Policy* 38 (2): 293–305.

de Moya-Anegón, F., Z. Chinchilla-Rodríguez, B. Vargas-Quesada, E. Corera-Álvarez, F. Munoz-Fernández, A. Gonzalez-Molina, and V. Herrero-Solana. 2007. "Coverage Analysis of Scopus: A Journal Metric Approach." *Scientometrics* 73 (1): 53–78.

de Wit, H., and F. Hunter. 2017. "Europe: The Future of Internationalization of Higher Education in Europe." In *Understanding Higher Education Internationalization. Insights from Key Global Publications*, edited by G. Mihut, P. G. Altbach, and H. de Wit, 25–28. Dordrecht: Sense.

Dobbins, M., and M. Kwiek. 2017. "Europeanisation and Globalisation in Higher Education in Central and Eastern Europe: 25 Years of Changes Revisited (1990–2015): Introduction to a Special Issue." *European Educational Research Journal* 16 (5): 519–28.

European Commission. 2007. *The European Research Area: New Perspectives*. Brussels: The European Commission.

European Commission. 2009. *Drivers of International Collaboration in Research*. Brussels: The European Commission.

European Commission. 2015. *Study on Network Analysis of the 7th Framework Programme Participation. Prepared by Science-Metrix, Fraunhofer ISI and Oxford Research*. Brussels: European Commission.

Finkelstein, M., and W. Sethi. 2014. "Patterns of Faculty Internationalization: A Predictive Model." In *The Internationalization of the Academy. Changes, Realities and Prospects*, edited by F. Huang, M. Finkelstein, and M. Rostan, 237–58. Dordrecht: Springer.

Finkelstein, M. J., E. Walker, and R. Chen. 2013. "The American Faculty in an Age of Globalization: Predictors of Internationalization of Research Content and Professional Networks." *Higher Education* 66 (3): 325–40.

Fox, M. F., M. L. Realff, D. R. Rueda, and J. Morn. 2017. "International Research Collaboration among Women Engineers: Frequency and Perceived Barriers, by Regions." *Journal of Technology Transfer* 42 (6): 1292–306.

Glänzel, W. 2001. "National Characteristics in International Scientific Co-authorship Relations." *Scientometrics* 51 (1): 69–115.

Glänzel, W., and A. Schubert. 2001. "Double Effort—Double Impact? A Critical View at International Co-authorship in Chemistry." *Scientometrics* 50 (2): 199–214.

Godin, B. 2007. "Science, Accounting and Statistics: The Input-Output Framework." *Research Policy* 36 (9): 1388–403.

Hennemann, S., and I. Liefner. 2015. "Global Science Collaboration." In *The Handbook of Global Science, Technology, and Innovation*, edited by D. Archibugi and A. Filippetti, 343–363. Somerset, NJ: Wiley.

Hoekman, J., K. Frenken, and R. J. Tijssen. 2010. "Research Collaboration at a Distance: Changing Spatial Patterns of Scientific Collaboration within Europe." *Research Policy* 41 (4): 520–31.

Jeong, S., J. Y. Choi, and J. Y. Kim. 2014. "On the Drivers of International Collaboration: The Impact of Informal Communication, Motivation, and Research Resources." *Science and Public Policy* 41 (4): 520–31.

Kamalski, J., and A. Plume. 2013. *Comparative Benchmarking of European and US Research Collaboration and Researchers Mobility: A Report Prepared in Collaboration Between Science Europe and Elsevier's SciVal Analytics*. Science Europe, Elsevier.

Kato, M., and A. Ando. 2017. "National Ties of International Scientific Collaboration and Researcher Mobility Found in Nature and Science." *Scientometrics* 110 (2): 673–94.

King, R. 2011. "Power and Networks in Worldwide Knowledge Coordination: The Case of Global Science." *Higher Education Policy* 24 (3): 359–76.

König, T. 2017. *The European Research Council*. Cambridge: Polity.

Kwiek, M. 2015. "The Internationalization of Research in Europe. A Quantitative Study of 11 National Systems from a Micro-Level Perspective." *Journal of Studies in International Education* 19 (2): 341–59.

Kwiek, M. 2016. "The European Research Elite: A Cross-National Study of Highly Productive Academics in 11 Countries." *Higher Education* 71 (3): 379–97.

Kwiek, M. 2018a. "International Research Collaboration and International Research Orientation: Comparative Findings about European Academics." *Journal of Studies in International Education* 22 (2): 136–60.

Kwiek, M. 2018b. "Academic Top Earners. Research Productivity, Prestige Generation and Salary Patterns in European Universities." *Science and Public Policy* 45 (1): 1–13.

Kwiek, M. 2018c. "High Research Productivity in Vertically Undifferentiated Higher Education Systems: Who Are the Top Performers?" *Scientometrics* 115 (1): 415–62.

Kwiek, M. 2019a. *Changing European Academics. A Comparative Study of Social Stratification, Work Patterns and Research Productivity*. London and New York: Routledge.

Kwiek, M. 2019b. "Social Stratification in Higher Education: What It Means at the Micro-Level of the Individual Academic Scientist." *Higher Education Quarterly* 73 (4): 419–44.

Kwiek, M. 2020. Internationalists and Locals: International Research Collaboration in a Resource-Poor System. *Scientometrics* Online first. doi:10.1007/s11192-020-03460-2.

Kwiek, M., and W. Roszka. 2020. "Gender Disparities in International Research Collaboration: A Large-Scale Bibliometric Study of 25,000 University Professors." *Journal of Economic Surveys* (submitted). https://arxiv.org/abs/2003.00537.

Kyvik, S., and D. W. Aksnes. 2015. "Explaining the Increase in Publication Productivity among Academic Staff: A Generational Perspective." *Studies in Higher Education* 40: 1438–53.

Lancho-Barrantes, B. S., V. P. Guerrero Bote, Z. C. Rodrigues, and F. de Moya Anegon. 2012. "Citation Flows in the Zones of Influence of Scientific Collaborations." *Journal of the American Society for Information Science and Technology* 63 (3): 481–9.

Larivière, V., E. Vignola-Gagné, C. Villeneuve, P. Gelinas, and Y. Gingras. 2011. "Sex Differences in Research Funding, Productivity and Impact: An Analysis of Quebec University Professors." *Scientometrics* 87 (3): 483–98.

Lasthiotakis, H., K. Sigurdson, and C. M. Sá. 2013. "Pursuing Scientific Excellence Globally: Internationalizing Research as a Policy Target." *Journal of Higher Education Policy and Management* 35 (6): 612–25.

Latour, B, and S Woolgar. 1986. *Laboratory life. The construction of scientific facts*. Princeton University Press.

Luukkonen, T., O. Persson, and G. Sivertsen. 1992. "Understanding Patterns of International Scientific Collaboration." *Science, Technology, & Human Values* 17 (1): 101–26.

Mattsson, P., P. Laget, A. Nilsson, and C.-J. Sundberg. 2008. "Intra-EU vs. Extra-EU Scientific Co-publication Patterns in EU." *Scientometrics* 75: 555–74.

Mayer, S. J., and J. M. Rathmann. 2018. "How Does Research Productivity Relate to Gender? Analyzing Gender Differences for Multiple Publication Dimensions." *Scientometrics* 117: 1663–93.

Melguizo, T., and M. H. Strober. 2007. "Faculty Salaries and the Maximization of Prestige." *Research in Higher Education* 48 (6): 633–68.

Olechnicka, A., A. Ploszaj, and D. Celinska-Janowicz. 2019. *The Geography of Scientific Collaboration*. London and New York: Routledge.

Payumo, J., T. Sutton, D. Brown, D. Nordquist, M. Evans, D. Moore, and P. Arasu. 2017. "Input-Output Analysis of International Research Collaboration: A Case Study of Five U.S. Universities. National Ties of International Scientific Collaboration and Research Mobility Found in Nature and Science." *Scientometrics* 111 (3): 1657–71.

Rodríguez-Navarro, A., and R. Brito. 2019. Might Europe One Day Again Be a Global Scientific Powerhouse? Analysis of ERC Publications Suggest It Will Not Be Possible Without Changes in Research Policy. Preprint, arXiv. https://arxiv.org/abs/1907.08975.

Royal Society. 2011. *Knowledge, Networks and Nations. Global Scientific Collaboration in the 21st Century*. London: The Royal Society.

Slaughter, S., and L. L. Leslie. 1997. *Academic Capitalism: Politics, Policies, and the Entrepreneurial University*. Baltimore: Johns Hopkins University Press.

Van den Besselaar, P., U. Sandström, and C. Mom. 2019. "Recognition through performance and reputation." In: Proceedings of the 17th International Conference on Scientometrics and Informetrics, ISSI 2019.

Wagner, C. S. 2006. "International Collaboration in Science and Technology: Promises and Pitfalls." In *Science and Technology Policy for Development, Dialogues at the Interface*, edited by L. Box, and R. Engelhard, 165–76. London: Anthem Press.

Wagner, C. S. 2008. *The new Invisible College. Science for Development*. Washington, DC: Brookings Institution Press.

Wagner, C. S. 2018. *The Collaborative Era in Science. Governing the Network*. Cham: Palgrave Macmillan.

Wagner, C. S., and L. Leydesdorff. 2005. "Network Structure, Self-Organization, and the Growth of International Collaboration in Science." *Research Policy* 34 (10): 1608–18.

Wagner, C. S., H. W. Park, and L. Leydesdorff. 2015. "The Continuing Growth of Global Cooperation Networks in Research: A Conundrum for National Governments." *PLoS ONE* 10 (7): 1–15.

Waltman, L., and N. J. van Eck. 2019. "Field Normalization of Scientometric Indicators." In *Springer Handbook of Science and Technology Indicators*, edited by W. Glänzel, H. F. Moed, U. Schmoch, and M. Thelwall, 281–300. Cham: Springer.

Whitley, R. 2000. *The Intellectual and Social Organization of the Sciences*. Oxford: Oxford University Press.

Woldegiyorgis, A. A., D. Proctor, and H. de Wit. 2018. "Internationalization of Research: Key Considerations and Concerns." *Journal of Studies in International Education* 22 (2): 161–76.

Yemini, M. 2019. "International Research Collaborations as Perceived by top-Performing Scholars." *Journal of Studies in International Education* 1–16. (online first November 9, 2019). doi:10.1177/1028315319887392 .